\documentclass[11pt,twoside,a4paper]{article}
\pdfoutput=1
\usepackage{jheppub}
\usepackage{bbm}
\usepackage{epsfig,multicol}

\usepackage{amsmath,amsthm,amssymb,bm,mathrsfs,float}
\usepackage[makeroom]{cancel}
\usepackage[hyperpageref]{backref} 
\usepackage{graphicx,epstopdf}

\title{Distribution theory for Schr{\"o}dinger's integral equation}
\author{Rutger-Jan Lange}
\emailAdd{rutger-jan.lange@cantab.net}

\abstract{\noindent Much of the literature on point interactions in quantum mechanics has focused on the differential form of Schr{\"o}dinger's equation. This paper, in contrast, investigates the integral form of Schr{\"o}dinger's equation. While both forms are known to be equivalent for smooth potentials, this is not true for distributional potentials. Here, we assume that the potential is given by a distribution defined on the space of discontinuous test functions. 

First, by using Schr{\"o}dinger's integral equation, we confirm a seminal result by Kurasov, which was originally obtained in the context of Schr{\"o}dinger's differential equation. This hints at a possible deeper connection between both forms of the equation. We also sketch a generalisation of Kurasov's result to hypersurfaces.

Second, we derive a new closed-form solution to Schr{\"o}dinger's integral equation with a delta prime potential. This potential has attracted considerable attention, including some controversy. Interestingly, the derived propagator  satisfies boundary conditions that were previously derived using Schr{\"o}dinger's differential equation.

Third, we derive boundary conditions for `super-singular' potentials given by higher-order derivatives of the delta potential. These boundary conditions cannot be incorporated into the normal framework of self-adjoint extensions. We show that the boundary conditions depend on the energy of the solution, and that probability is conserved. 

This paper thereby confirms several seminal results and derives some new ones. In sum, it shows that Schr{\"o}dinger's integral equation is viable tool for studying singular interactions in quantum mechanics.

}
\keywords{point interaction, self-adjoint extension (SAE), singular potential, delta potential, delta prime potential, surface delta function, surface delta prime function, distribution theory, discontinuous test function}

\begin{document}
\pagestyle{myplain}
\maketitle


\setcounter{section}{0}
\setcounter{subsection}{0}
\setcounter{equation}{0}
\setcounter{figure}{0}


\numberwithin{equation}{section}


\section{Introduction}
\label{section1}

It has long been known that the Dirac delta potential allows for an exact solution to the time-dependent Schr{\"o}dinger equation. Equally well known are the corresponding boundary conditions. It may be surprising, therefore, that the Dirac delta \textit{prime} potential has caused headaches, and the corresponding boundary conditions have been subject to debate for much of the last three decades (see e.g. \cite{Albeverio1984,Seba1986,Seba1986b,Gesztesy1987,Albeverio1988,Zhao1992,Albeverio1993,Griffiths1993,Carreau1993,Patil1994,Albeverio1994,Grosche1995,Nogami1996scattering,Park1996,Roman1996,Coutinho1997,Albeverio1998,Cheon1998,Coutinho1999,coutinho1999many,Griffiths1999,Albeverio2000singular,christiansen2003existence,Coutinho2005,zolotaryuk2006scattering,Toyama2007,zolotaryuk2007resonant,Araujo2008time,zolotaryuk2008two,coutinho2009unusual,gadella2009bound,golovaty2009solvable,golovaty2010norm,zolotaryuk2010boundary,zolotaryuk2010point,coutinho2012one,golovaty2012schroedinger,lange2012,albeverio2013remarkable,brasche2013one,Golovaty2013,zolotaryuk2013,zolotaryuk2014,Lunardi2014}). It is worth discussing some of the ambiguities surrounding the delta prime potential in more detail (see also Table \ref{Table1}):

\begin{itemize}
\item \textbf{Ambiguous Schr{\"o}dinger equation}: It has been assumed (correctly) that the wave function $\psi$ is discontinuous in the presence of a delta prime potential. However, the Schr{\"o}dinger equation is then ambiguous (see e.g. \cite{kurasov1996distribution}). For many constructions of the delta prime, e.g. methods 2-4 in Table 1, the integral $\int \delta' \psi$ blows up, since the `slope'  of $\psi$ is infinite at the origin.

\item \textbf{Arbitrary boundary conditions:} To resolve this issue, many authors have decided that the delta prime potential is not to be taken literally. Instead, they define the delta prime \textit{interaction} (as opposed to the delta prime potential) by some self-adjoint boundary condition. A jump in the value but not in the derivative is often assumed \cite{Gesztesy1985,Seba1986,Gesztesy1987,Albeverio1993,Albeverio2005,Behrndt2013,zolotaryuk2014}. However, this assumption is arbitrary at best and misleading at worst, as pointed out in \cite{Exner1996,Coutinho1997,christiansen2003existence}. 

\item \textbf{Ambiguous limits}: Several authors have explicitly solved Schr{\"o}dinger's differential equation for potentials which, in the limit, are equal to the delta prime function. The boundary conditions can then be read off. The transition and reflection properties, however, depend crucially on `hidden parameters' that determine how the potential approaches the limit (see e.g. \cite{Toyama2007,golovaty2010norm,zolotaryuk2010boundary,zolotaryuk2010point,christiansen2003existence}). Further, this approach does not in general resolve the ambiguity of the Schr{\"o}dinger equation, in the sense that $\int \delta' \psi$ does not generally exist if $\psi$ is disontinuous.
\end{itemize}
Our approach is different in that we investigate the \textit{integral form} of Schr{\"o}dinger's equation. We assume that the potential is equal to some distribution defined on the space of discontinuous test functions. 

First, we replicate a seminal result by Kurasov  \cite{kurasov1996distribution}, which is based on distribution theory for the \textit{differential form} of Schr{\"o}dinger's equation. This is both reassuring and somewhat surprising, since the equivalence of both approaches is guaranteed only for smooth potentials. Our result thus hints at a deeper connection between the integral and differential forms of Schr{\"o}dinger's equation. 

\begin{table}[b]
\begin{tabular}
{|l|c|c|l|}
\hline
\textbf{Method} 
& \textbf{Literature}
& \textbf{Definition}																			
& \textbf{Drawback} 
\\[2ex]\hline

1. `Label'
&\cite{Albeverio1984,Seba1986b,Gesztesy1987,Albeverio1993,Albeverio1994}
& $\psi'(0^{+})=\psi'(0^{-})$
& ABC
\\

for some BCs
&\cite{Coutinho1997,Albeverio1998,Coutinho1999,coutinho1999many,Albeverio2000singular}
& $\psi(0^{+})-\psi(0^{-}) \propto \psi'(0) $											
& 
\\\hline

2. Dipole interaction	
&\cite{Seba1986,Patil1994,coutinho2009unusual}						
& $\displaystyle\underset{\epsilon \searrow 0}\lim\;\frac{1}{\epsilon^\nu}\; [\delta(x+\epsilon)+\delta(x-\epsilon)]$		
& ASE																	
\\\hline

3. Rectangular  	
&\cite{christiansen2003existence,zolotaryuk2006scattering,zolotaryuk2008two}
& $\displaystyle\underset{\epsilon,l\searrow 0}\lim\;\frac{1}{\epsilon\,l}\;\Big[\mathbbm{1}_{[\frac{-l-\epsilon}{2}<x<\frac{-l+\epsilon}{2}]}\ldots$
&	ASE, AL
\\

approximation 
& 
& $\ldots-\mathbbm{1}_{[\frac{l-\epsilon}{2}<x<\frac{l+\epsilon}{2}]}\Big]$  	
&																													
\\\hline

4. Short-range	
& \cite{golovaty2009solvable,golovaty2010norm}
& $\displaystyle \underset{\epsilon \searrow 0}\lim\,\frac{1}{\epsilon^2} V(x/\epsilon)$
& ASE
\\

potentials	
& \cite{golovaty2012schroedinger,Golovaty2013}
& s.t. $\int V =0, \int x V =-1$ 
& 
\\\hline
\end{tabular}
 \caption{Overview of common definitions of the delta prime potential in the literature. Possible drawbacks are an ambiguous Schr{\"o}dinger equation (ASE), arbitrary boundary conditions (ABC), and ambiguous limits (AL).}
  \label{Table1}  	
\end{table} 

Second, we consider Schr{\"o}dinger's integral equation with a delta prime potential. As pointed out above, this potential has attracted considerable interest in the literature. We derive a new and exact solution for the time-dependent propagator. This solution satisfies boundary conditions previously derived by some authors in the context of distribution theory for Schr{\"o}dinger's differential equation, thereby further emphasizing the apparent equivalence of both approaches.

Third, we use Schr{\"o}dinger's integral equation to derive boundary conditions for higher-order derivatives of the delta potential. Such `super-singular' potentials are of interest as they cannot be incorporated into the usual framework of self-adjoint extensions. We find that the associated boundary conditions are of the self-adjoint form --- but with the crucial difference that the constants in the boundary conditions depend on the \textit{energy} of the solution. We show that probability is conserved for these engery-dependent point interactions. 

This paper is structured as follows. Section \ref{section2} re-derives Kurasov's potential based purely on a symmetry argument. Section \ref{section3} re-writes the corresponding boundary conditions concisely in the \textit{jump-average} form. Section \ref{section4} extends Kurasov's result by showing that these boundary conditions follow directly from Schr{\"o}dinger's \textit{integral} equation. Section \ref{section5} proposes to further extend this result to hypersurfaces. Section \ref{section6} presents the scattering matrix in one dimension. Sections \ref{section7} and \ref{section8} show that the jump-average boundary conditions form a subset of all possible self-adjoint extensions. Section \ref{section9} derives a new, exact result for the propagator in the presence of a delta prime potential. Section \ref{section10} show that super-singular potentials, given by higher-order derivatives of the delta function, lead to energy-dependent boundary conditions that conserve probability. Section \ref{section11}, finally, sums up our findings and points to future research.

\section{Kurasov's potential revisited}
\label{section2}

Suppose we seek a Hermitian operator that connects the Dirac delta function with a maximum of two differential operators. We quickly see that we can construct \textit{three} fundamental point interactions, namely
\begin{equation}
\label{potential}
\begin{array}{rcl}
V(x)&=&\displaystyle c _1\,\delta(x) + c _2\,\frac{d}{dx}\,\delta(x) - \overline{c }_2\,\delta(x)\,\frac{d}{dx} + c _3\, \frac{d}{dx}\,\delta(x)\,\frac{d}{dx}.
\end{array}
\end{equation}
It is understood that differential operators differentiate everything to their right. Complex conjugation is denoted by $\overline\cdot$. The requirement that $V$ is Hermitian implies $c_1,c_3 \in \mathbbm{R}$, while $c_2 \in \mathbbm{C}$ is allowed. The action of the Dirac delta function on possibly discontinuous test functions has not yet been defined. The maximal domain of this operator is the Sobolev space $W_2^2(\mathbbm{R}\backslash 0)$. 

Assume the Dirac delta function is even under parity. Then it holds that the first and third point interactions, defined by $c_1$ and $c_3$, are also even, since they contain an even number of derivatives. The second point interaction, defined by $c_2$ and $\overline{c}_2$, on the other hand, is odd. If $c _2$ is real, the potential simplifies to
\begin{equation*}
\begin{array}{rcl}
V&=&\displaystyle  c _1\,\delta(x) + c _2\,\delta'(x) + c _3\, \frac{d}{dx}\,\delta(x)\,\frac{d}{dx}.
\end{array}
\end{equation*}
The operator \eqref{potential} was discovered in \cite{kurasov1996distribution} by an entirely different route. We can make the correspondence explicit by taking
\begin{equation*}
c _1 = X_1,\quad c _2 = X_2 + \mathbbm{i} X_3,\quad c _3 =-X_4,\quad\quad X_1,X_2,X_3,X_4 \in \mathbbm{R}.
\end{equation*} 
In that notation, $L=-d^2/dx^2+V$ can be written as
\begin{equation}
\label{kurasovL}
\begin{array}{rcl}
L&=&\displaystyle-\frac{d^2}{dx^2}+ X_1\,\delta(x) + \mathbbm{i} \frac{d}{dx} \Big(2 X_3 \delta(x) - \mathbbm{i} X_4 \delta'(x)\Big)+\Big(X_2-\mathbbm{i}X_3\Big)\delta'(x) - X_4 \frac{d^2}{d x^2}\delta(x),
\end{array}
\end{equation}
which corresponds exactly to \cite[p. 307]{kurasov1996distribution}. Our representation, which is different only in form, further underpins Kurasov's operator by showing that it follows directly from symmetry considerations. 

More recently, \cite{AlbeverioKuzhel2005,BrascheNizhnik2013,CojuhariGrodKuzhel2014} have also considered operators of the form \eqref{potential}. In \cite{AlbeverioKuzhel2005} and \cite{CojuhariGrodKuzhel2014}, four independent complex numbers were allowed in place of our $c_1$, $c_2$, $\overline{c}_2$ and $c_3$. In \cite[p. 4978]{AlbeverioKuzhel2005}, the form \eqref{potential} was subsequently derived using symmetry considerations.

The historical labels associated with these point interactions are summarised in \cite{BrascheNizhnik2013}, although it must be said that they can be somewhat misleading. Instead, we will simply refer to interactions defined by $c_1$, $c_2$ and $c_3$ as the first, second and third fundamental point interactions.  

\section{Jump-average boundary conditions}
\label{section3}

The boundary conditions corresponding to the operator \eqref{kurasovL} were derived in the context of Schr{\"o}dinger's differential equation in \cite[p. 307-308]{kurasov1996distribution}. As it turns out, however, the resulting boundary conditions can be expressed differently, and quite naturally, using the average and discontinuity of the solution. To this end, we define $\{u\}$ and $[u]$ as follows:
\begin{equation*}
\begin{array}{crl}
\{u(0)\}=\displaystyle\frac{u(0^+)+u(0^-)}{2},\quad\quad\quad\quad [u(0)]=u(0^+)-u(0^-).
\end{array}
\end{equation*}
As in \cite{kurasov1996distribution}, we suppose that the action of the Dirac delta function on the space of discontinuous functions $u \in W_2^{(n+1)}(\mathbbm{R}\backslash 0)$ is defined by
\begin{equation}
\label{Dirac1}
\int_{-\infty}^{\infty} \delta^{(n)}(x)\,u(x)\,dx= (-1)^n\{u^{(n)}(0)\}.
\end{equation}
The boundary conditions associated with the operator \eqref{kurasovL}, can now be written in compact form as
\begin{equation}
\label{SAE6}
\left(
\begin{array}{c}
[u'(0)]\\
{[u(0)]}
\end{array}
\right)
= 
\left(
\begin{array}{c@{\hspace{7mm}}c}
c _1& -\overline{c }_2\\
c _2 & c _3
\end{array}
\right)
\left(
\begin{array}{c}
\{u(0)\}\\
\{u'(0)\}
\end{array}
\right), \quad c_1, c_3 \in \mathbbm{R},\,c_2 \in \mathbbm{C}.
\end{equation}
We will refer to these boundary conditions as the \textit{jump-average boundary conditions}. In appearance they are quite different to the boundary conditions originally derived in \cite[p. 307-308]{kurasov1996distribution}, but they are identical in content. In the next section, we will re-derive this important result --- but in the context of the \textit{integral} form of Schr{\"o}dinger's equation. 

Boundary conditions of the jump-average form first seem to have appeared in \cite{AlbeverioNizhnik2006}. More recently, they were used in \cite{BrascheNizhnik2013,CojuhariGrodKuzhel2014}. There, it was initially supposed that an \textit{arbitrary} complex matrix connects the jumps to the averages --- thus allowing eight degrees of freedom. A different set of papers has considered jump-average boundary conditions with the  additional (but unnecessary) requirement that $c_2$ is real (see \cite{Wu2002,Coutinho2004,Coutinho2005,coutinho2012one}). 

An attractive property of the jump-average representation, which seems to have been overlooked in the literature, is its behaviour under parity. As $x \to -x$, we get
\begin{equation*}
\left(
\begin{array}{c}
[u'(0)]\\
{[u(0)]}
\end{array}
\right)
\to 
P
\left(
\begin{array}{c}
[u'(0)]\\
{[u(0)]}
\end{array}
\right),\quad 
\left(
\begin{array}{c}
\{u(0)\}\\
\{u'(0)\}
\end{array}
\right)
\to 
P
\left(
\begin{array}{c}
\{u(0)\}\\
\{u'(0)\}
\end{array}
\right),\quad\mbox{where}\;
P=\left(
\begin{array}{c@{\hspace{5mm}}c}
1& 0\\
0& -1
\end{array}
\right).
\end{equation*}
As a result, the connection matrix changes to
\begin{equation*}
\left(
\begin{array}{c@{\hspace{7mm}}c}
c _1& -\overline{c }_2\\
c _2 & c _3
\end{array}
\right)
\to 
P
\left(
\begin{array}{c@{\hspace{7mm}}c}
c _1& -\overline{c }_2\\
c _2 & c _3
\end{array}
\right)
P^{-1}
= 
\left(
\begin{array}{c@{\hspace{7mm}}c}
c _1& \overline{c }_2\\
-c _2 & c _3
\end{array}
\right).
\end{equation*}
Thus $c_1$ and $c_3$ are even under parity, while $c_2$ is odd. Indeed, this was to be expected from the heuristic reasoning which led to the potential \eqref{potential}. For future reference, we note that the determinant $D=c_1\,c_3 + |c_2|^2$ is real and even under parity. 

\section{Integral equation with Kurasov's potential}
\label{section4}

This section considers Schr{\"o}dinger's \textit{integral} equation with the potential \eqref{potential}, which reads (see e.g. \cite{FeynmanHibbs1965,Schulman1981techniques,Gaveau1986,Park1996}):
\begin{equation}
\label{SchrodingerEquation4}
\begin{array}{c}
\displaystyle\psi(y,t|x,s)=\displaystyle \psi_0(y,t|x,s)-\mathbbm{i} \,\int_s^t d\tau \int^\infty_{-\infty} d\alpha\;\psi_0(y,t|\alpha,\tau)\,V(\alpha)\,\psi(\alpha ,\tau |x,s).\\
\end{array}
\end{equation}
As in \cite{kurasov1996distribution}, we take $\hbar=1$ and $m=1/2$. In these units, the free propagator $\psi_0$ reads
\begin{equation}
\label{freepropagator}
\psi_0(y,t|x,s)=\frac{1}{\sqrt{4\pi\, \mathbbm{i}\, (t-s)}} \exp \Big[ \frac{-(y-x)^2}{4\,\mathbbm{i}\,(t-s)}\Big ],\quad\quad t>s.
\end{equation}
If the potential is singular, then $\psi$ is not generally continuous. It is crucial, therefore, to define the potential as a distribution acting on the space of discontinuous test functions; otherwise the integral equation \eqref{SchrodingerEquation4} goes undefined. For example, it is tempting to define the delta function as the limit of a Gaussian, and the delta prime as the limit of the derivative of a Gaussian. But then the integral equation \eqref{SchrodingerEquation4} with the potential \eqref{potential} has no solution. In that case, $\int \delta' \psi$ blows up for discontinuous $\psi$. Since the integral equation does not \textit{allow} continuous solutions, and does not \textit{exist} for discontinuous solutions, it has no solutions at all. In fact, only for a definition of the delta function (and its derivatives) that allows discontinuous test functions is there a solution to  Schr{\"o}dinger's integral equation.

The smoothness assumptions required on $\psi(\cdot,t|x,s)$ depend on the singularity of the potential. For the potential \eqref{potential}, it is sufficient to assume $\psi(\cdot,t|x,s)\in W_2^2(\mathbbm{R}\backslash\{0\})$. Then Schr{\"o}dinger's integral equation reads:
\begin{equation*}
\begin{array}{r@{\hspace{1mm}}c@{\hspace{1mm}}ll}
\psi(y,t|x,s)&=&\displaystyle \psi_0(y,t|x,s) & \displaystyle -\,c _1\,\mathbbm{i}\displaystyle\int_s^t d\tau \int_{-\infty}^{\infty}d\alpha\;\psi_0(y,t|\alpha,\tau)\,\delta(\alpha)\,\psi(\alpha,\tau|x,s)\\[2ex]
&&& \displaystyle -\,c _2\,\mathbbm{i}\displaystyle\int_s^t d\tau\int_{-\infty}^{\infty}d\alpha\;\psi_0(y,t|\alpha,\tau)\,\frac{d}{d\alpha}\Bigg(\delta(\alpha)\,\psi(\alpha,\tau|x,s)\Bigg)\\[2ex]
&&& \displaystyle +\,\overline{c }_2\,\mathbbm{i}\displaystyle\int_s^t d\tau\int_{-\infty}^{\infty}d\alpha\;\psi_0(y,t|\alpha,\tau)\,\delta(\alpha)\,\frac{d}{d\alpha}\psi(\alpha,\tau|x,s)\\[2ex]
&&& \displaystyle -\,c _3\,\mathbbm{i}\displaystyle\int_s^t d\tau\int_{-\infty}^{\infty}d\alpha\;\psi_0(y,t|\alpha,\tau)\,\frac{d}{d\alpha}\Bigg(\delta(\alpha)\,\frac{d}{d\alpha}\,\psi(\alpha,\tau|x,s)\Bigg).
\end{array}
\end{equation*}
The manipulations that follow are relatively straightforward. First, by writing out all differentiations, we obtain
\begin{equation*}
\begin{array}{r@{\hspace{1mm}}c@{\hspace{1mm}}ll}
\psi(y,t|x,s)&=&\displaystyle \psi_0(y,t|x,s) & \displaystyle -\,c _1\,\mathbbm{i}\displaystyle\int_s^t d\tau \int_{-\infty}^{\infty}d\alpha\;\psi_0(y,t|\alpha,\tau)\,\delta(\alpha)\,\psi(\alpha,\tau|x,s)\\[2ex]
&&& \displaystyle -\,c _2\,\mathbbm{i}\displaystyle\int_s^t d\tau\int_{-\infty}^{\infty}d\alpha\;\psi_0(y,t|\alpha,\tau)\,\delta'(\alpha)\,\psi(\alpha,\tau|x,s)\\[2ex]
&&&\displaystyle -\,c _2\,\mathbbm{i}\displaystyle\int_s^t d\tau\int_{-\infty}^{\infty}d\alpha\;\psi_0(y,t|\alpha,\tau)\,\delta(\alpha)\,\psi'(\alpha,\tau|x,s)\\[2ex]
&&& \displaystyle +\,\overline{c }_2\,\mathbbm{i}\displaystyle\int_s^t d\tau\int_{-\infty}^{\infty}d\alpha\;\psi_0(y,t|\alpha,\tau)\,\delta(\alpha)\,\psi'(\alpha,\tau|x,s)\\[2ex]
&&& \displaystyle -\,c _3\,\mathbbm{i}\displaystyle\int_s^t d\tau\int_{-\infty}^{\infty}d\alpha\;\psi_0(y,t|\alpha,\tau)\,\delta'(\alpha)\, \psi'(\alpha,\tau|x,s)\\[2ex]
&&& \displaystyle -\,c _3\,\mathbbm{i}\displaystyle\int_s^t d\tau\int_{-\infty}^{\infty}d\alpha\;\psi_0(y,t|\alpha,\tau)\,\delta(\alpha)\,\psi''(\alpha,\tau|x,s).
\end{array}
\end{equation*}
Primes denote differentiation with respect to $\alpha$. Second, using the definition of the Dirac delta function in \eqref{Dirac1}, we get
\begin{equation*}
\begin{array}{r@{\hspace{1mm}}c@{\hspace{1mm}}ll}
\psi(y,t|x,s)&=&\displaystyle \psi_0(y,t|x,s) & \displaystyle -\,c _1\,\mathbbm{i}\displaystyle\int_s^t d\tau \;\psi_0(y,t|0,\tau)\,\{\psi(0,\tau|x,s)\}\\[2ex]
&&& \displaystyle +\,c _2\,\mathbbm{i}\displaystyle\int_s^t d\tau\;\Big\{\psi_0'(y,t|0,\tau)\psi(0,\tau|x,s)+\psi_0(y,t|0,\tau)\psi'(0,\tau|x,s)\Big\}\\[2ex]
&&&\displaystyle -\,c _2\,\mathbbm{i}\displaystyle\int_s^t d\tau\;\psi_0(y,t|0,\tau)\,\{\psi'(0,\tau|x,s)\}\\[2ex]
&&&\displaystyle +\,\overline{c} _2\,\mathbbm{i}\displaystyle\int_s^t d\tau\;\psi_0(y,t|0,\tau)\,\{\psi'(0,\tau|x,s)\}\\[2ex]
&&& \displaystyle +\,c _3\,\mathbbm{i}\displaystyle\int_s^t d\tau\;\Big\{\psi_0'(y,t|0,\tau)\,\psi'(0,\tau|x,s)+\psi_0(y,t|0,\tau)\,\psi''(0,\tau|x,s)\Big\}\\[2ex]
&&& \displaystyle -\,c _3\,\mathbbm{i}\displaystyle\int_s^t d\tau\;\psi_0(y,t|0,\tau)\,\{\psi''(0,\tau|x,s)\}.
\end{array}
\end{equation*}
Since the free propagator $\psi_0$ is smooth, it can be pulled out of the averaging operator. Four terms (two pairs) cancel, and we obtain
\begin{equation*}
\begin{array}{r@{\hspace{1mm}}c@{\hspace{1mm}}ll}
\psi(y,t|x,s)&=&\displaystyle \psi_0(y,t|x,s) & \displaystyle -\,c _1\,\mathbbm{i}\displaystyle\int_s^t d\tau \;\psi_0(y,t|0,\tau)\,\{\psi(0,\tau|x,s)\}\\[2ex]
&&& \displaystyle +\,c _2\,\mathbbm{i}\displaystyle\int_s^t d\tau\;\psi_0'(y,t|0,\tau)\{\psi(0,\tau|x,s)\}\\[2ex]
&&&\displaystyle +\,\overline{c} _2\,\mathbbm{i}\displaystyle\int_s^t d\tau\;\psi_0(y,t|0,\tau)\,\{\psi'(0,\tau|x,s)\}\\[2ex]
&&& \displaystyle +\,c _3\,\mathbbm{i}\displaystyle\int_s^t d\tau\;\psi_0'(y,t|0,\tau)\,\{\psi'(0,\tau|x,s)\}.
\end{array}
\end{equation*}
Finally, the free propagator $\psi_0(y,t|\alpha,\tau)$ satisfies $\partial_\alpha \psi_0 = -\partial_y \psi_0$. Therefore 
\begin{equation}
\label{integralequation}
\begin{array}{lrl}
\psi(y,t|x,s)&=\displaystyle \psi_0(y,t|x,s) - \displaystyle c _1\,\mathbbm{i}&\displaystyle\int_s^t  \;\psi_0(y,t|0,\tau)\,\{\psi(0,\tau|x,s)\}\,d\tau\\[2ex]
&- \displaystyle c _2\,\mathbbm{i}\,\frac{d}{dy}\,&\displaystyle\int_s^t \;\psi_0(y,t|0,\tau)\,\{\psi(0,\tau|x,s)\}\,d\tau\\[2ex]
&+ \displaystyle \overline{c }_2\,\mathbbm{i}&\displaystyle\int_s^t \;\psi_0(y,t|0,\tau)\,\{\psi'(0,\tau|x,s)\}\,d\tau\\[2ex]
&- \displaystyle c _3\,\mathbbm{i}\,\frac{d}{dy}\,&\displaystyle\int_s^t \;\psi_0(y,t|0,\tau)\,\{\psi'(0,\tau|x,s)\}\,d\tau.
\end{array}
\end{equation}
The derivatives with respect to $y$ have been pulled to the outside of the integrals. This is allowed for all $y\neq 0$, where $\psi(\cdot,t|x,s)$ is smooth. As a result, we can meaningfully speak of $\{\psi(0,t|x,s)\}$ and $[\psi(0,t|x,s)]$. Of course, the quantities $\psi(0,t|x,s)$ and $\psi'(0,t|x,s)$ have no meaning.

The jump-average boundary conditions follow directly from the integral equation \eqref{integralequation}. To see why, consider the auxiliary function $f$, defined as 
\begin{equation}
\label{fdefinition}
\begin{array}{lr@{\hspace{1mm}}c@{\hspace{1mm}}}
f(y,t|x,s):=\displaystyle -\mathbbm{i}\int_s^t \; \psi_0(y,t|0,\tau)\,g(\tau|x,s)\,d\tau,
\end{array}
\end{equation}
where $g$ is some other function. Note that all integral terms on the right-hand side of \eqref{integralequation} can be written as either $f$ or as $\partial_y f$ for some $g$. It can be shown that $f(\cdot,t|x,s)$ is discontinuous only for odd derivatives. Specifically,
\begin{equation}
\label{fconditions}
\begin{array}{r@{\hspace{1mm}}c@{\hspace{1mm}}l}
\displaystyle [f(0,t|x,s)]&=&0,\\[2ex]
\displaystyle [f^{(1)}(0,t|x,s)]&=&\displaystyle\,g(t|x,s),\\[2ex]
\displaystyle [f^{(2)}(0,t|x,s)]&=&0.\\[2ex]
\end{array}
\end{equation} 
This implies that $[\psi(0,t|x,s)]$ is determined purely by the second and fourth integrals in \eqref{integralequation}, which have the derivative $d/dy$ in front of them. Similarly, $[\psi'(0,t|x,s)]$ is determined purely by the first and third integrals in \eqref{integralequation}, which have no derivative. The solution $\psi$, which appears on the left-hand side, inherits the  discontinuities of all terms on the right-hand side. Thus, by \eqref{fconditions}, the integral equation \eqref{integralequation} implies 
\begin{equation}
\label{derivedBCs}
\left(
\begin{array}{c}
[\psi'(0,t|x,s)]\\
{[\psi(0,t|x,s)]}
\end{array}
\right)
= 
\left(
\begin{array}{c@{\hspace{7mm}}c}
c _1& -\overline{c }_2\\
c _2 & c _3
\end{array}
\right)
\left(
\begin{array}{c}
\{\psi(0,t|x,s)\}\\
\{\psi'(0,t|x,s)\}
\end{array}
\right).
\end{equation}
Thus Schr{\"o}dinger's integral equation with the potential \eqref{potential} implies the jump-average boundary conditions \eqref{SAE6}. While our conclusion is consistent with \cite{kurasov1996distribution} this was not a priori obvious, given that the differential and integral forms of Schr{\"o}dingers equation are known to be equivalent only for smooth potentials. Our result thus hints at a possible deeper connection between both forms of Schr{\"o}dinger's equation. A further advantage of our method is that it can be extended relatively easily to hypersurfaces (see the next section) and to super-singular potentials (see section \ref{section10}).

\section{Extension to hypersurfaces}
\label{section5}

This section sketches informally how Kurasov's result, as re-derived in the previous section, may be generalised to surfaces of co-dimension one. A rigorous treatment would define self-adjoint operators acting on Sobolev spaces, and show resolvent convergence of operators used to approximate singular potentials. For the sake of brevity, however, we will confine ourselves to a heuristic treatment only. It is hoped that the reader will permit this brief digression, which demonstrates, albeit not overly rigorously, a neat link with classical potential theory.

As is well known, Dirac's delta function can be defined (purely formally) as the derivative of the Heaviside step function. In other words: as the inward-pointing derivative of the indicator function of the positive halfline. In higher dimensions, we argue, it is natural to consider the inward normal derivative of the indicator function of some domain $D$. 

Let $S$ be a smooth hypersurface enclosing some domain $D$ in $d$ dimensions, where the \textit{inside} of $S$ is defined to be the side where $D$ is located. As in \cite{lange2012}, we define the surface delta function as $\delta_S(x)=n_x \cdot \nabla_x \mathbbm{1}_{x \in D}$, where $n_x$ is the inward normal, $\nabla_x$ is the gradient operator, and $\mathbbm{1}_{x \in D}$ is the indicator function of the domain $D$. Similarly, again as in  \cite{lange2012}, we define the surface delta prime function $\delta'_S(x)=n_x \cdot \nabla_x \delta_S(x) = \nabla_x^2\mathbbm{1}_{x \in D}$, i.e. as the Laplacian of the indicator function. Then, we extend these definitions to allow for discontinuous test functions as follows:
\begin{equation}
\label{Diracsurface}
\begin{array}{r@{\hspace{1mm}}c@{\hspace{1mm}}l}
\displaystyle \underset{\mathbb{R}^d}\int\delta_S(x)\,u(x)\;dx&=&\displaystyle \underset{S}\int \; \{u(\beta)\}\;d\beta,\\[2ex]
\displaystyle \underset{\mathbb{R}^d}\int\delta'_S(x)\,u(x)\;dx&=&-\displaystyle \underset{S}\int\; \{ u'(\beta)\} \;d\beta.
\end{array}
\end{equation}
In analogy with one dimension, we use $\{ \cdot \}$ and $[ \cdot ]$ to denote the average and discontinuity across the surface $S$ in the inward normal direction, while a prime denotes the normal derivative, also in the inward direction\footnote{By taking $D$ to be the positive real line, the one dimensional formulas are recovered.}. With these definitions, we propose the following hypersurface generalisation of \eqref{potential}:
\begin{equation}
\label{potentiald}
\begin{array}{rcl}
V(x)&=&\displaystyle c _1\,\delta_S(x) + c _2\left(n_x\cdot \nabla_x\right)\,\delta_S(x) - \overline{c }_2\,\delta_S(x)\,\left(n_x\cdot \nabla_x\right) + c _3\left(n_x\cdot \nabla_x\right)\,\delta_S(x)\,\left(n_x\cdot\nabla_x\right). 
\end{array}
\end{equation}
In \eqref{potential}, we have simply replaced $\delta(x)$ by $\delta_S(x)$ and $d/dx$ by $n_x \cdot \nabla_x$. As in one dimension, we have $c _1,c _3 \in \mathbbm{R}$ and $c _2 \in \mathbbm{C}$. If $c_2$ is real, the potential simplifies to
\begin{equation*}
\begin{array}{rcl}
V(x)&=&\displaystyle c _1\,\delta_S(x) + c _2\, \delta_S'(x) + c _3\left(n_x\cdot \nabla_x\right)\,\delta_S(x)\,\left(n_x\cdot\nabla_x\right).
\end{array}
\end{equation*}
To complete our problem set-up, we consider a wave-function  $\psi(\cdot,\cdot|x,s)$ that satisfies Schr{\"o}dinger's integral equation in $\mathbbm{R}^d$, i.e.
\begin{equation}
\label{SchrodingerEquationd}
\makebox[13cm][l]
{
$
\begin{array}{c}
\displaystyle\psi(y,t|x,s)=\displaystyle \psi_0(y,t|x,s)-\mathbbm{i} \,\int_s^t d\tau \underset{\mathbbm{R}^d}{\int} d\alpha\;\psi_0(y,t|\alpha,\tau)\,V(\alpha)\,\psi(\alpha ,\tau |x,s),\\
\end{array}
$
}
\end{equation}
where the potential $V$ is given by \eqref{potentiald}, and $\psi_0$ now equals the free propagator in $d$ dimensions, with the usual conventions that $\hbar=1$ and $m=1/2$. 

By the same approach as in one dimension, Schr{\"o}dinger's integral equation \eqref{SchrodingerEquationd} with the potential $V$ as in \eqref{potentiald} implies that $\psi(\cdot,t|x,s)$ must satisfy the following \textit{surface jump-average} boundary conditions:
\begin{equation}
\label{SAE6d}
\left(
\begin{array}{c}
[\psi'(\beta,t|x,s)]\\
{[\psi(\beta,t|x,s)]}
\end{array}
\right)
= 
\left(
\begin{array}{c@{\hspace{7mm}}c}
c _1& -\overline{c }_2\\
c _2 & c _3
\end{array}
\right)
\left(
\begin{array}{c}
\{\psi(\beta,t|x,s)\}\\
\{\psi'(\beta,t|x,s)\}
\end{array}
\right),\quad \forall \beta \in S.
\end{equation}
These boundary conditions form a self-adjoint extension of the Laplacian, and probability is conserved locally, i.e. for each point on the surface. These boundary conditions have some interesting implications. It can be verified directly that $c_2=2$ with $c_1=c_3=0$ leads to Neumann boundary conditions on the \textit{inside} of $S$, and Dirichlet boundary conditions on the \textit{outside} of $S$. Conversely, $c_2=-2$ (again with $c_1=c_3=0$) leads to Dirichlet boundary conditions on the \textit{inside} of $S$, and Neumann boundary conditions on the \textit{outside} of $S$.

By a rotation to imaginary time, i.e $t \to -\mathbbm{i}\,t$, the free propagator $\psi_0$ turns into the propagation density of a $d$-dimensional Brownian motion. The propagator of a Brownian motion started in the interior of $D$ and absorbed (reflected) on the surface $S$ satisfies Dirichlet (Neumann) boundary conditions there. Focusing on the inside of $S$, it turns out that these boundary conditions are generated by $c_2=-2$ ($c_2=+2$), i.e. by the surface delta prime potential $V(x)=\mp 2 \delta'_S(x)$. Intriguingly, as first noted in \cite{lange2012}, the only difference between the classical Dirichlet/Neumann boundary value problems for the Brownian propagator resides in the \textit{sign} of the potential! 

Finally, Robin boundary conditions on the inside of $S$ are generated by $c_2=2$, $c_3=0$ and $c_1$ being real and non-zero. As noted, these results are not overly rigorous; however, this section has demonstrated that interactions on surfaces of co-dimension one are a natural generalisation of point interactions in one dimension.

\section{Scattering matrix in one dimension}
\label{section6}

This section presents the scattering coefficients for the three fundamental point interactions in one dimension. Although the result is straightforward to obtain, it is quite insightful. Consider a stationary wave $\psi_+$ incoming from the left and moving towards the right, i.e.
\begin{equation}
\label{stationarystate}
\psi_+(x)=
\left\{
\begin{array}{l@{\hspace{15mm}}l}
e^{\mathbbm{i}\,k\,x}+R_+\,e^{-\mathbbm{i}\,k\,x},&x<0,\\
T_+\,e^{\mathbbm{i}\,k\,x},&x>0,
\end{array}
\right.
\end{equation}
where $k$ is related to the energy by $k^2=E$. Similarly, we denote by $\psi_-$ a stationary wave that is moving towards the left with transmission and reflection coefficients $T_-$ and $R_-$. Imposing the jump-average boundary conditions \eqref{SAE6} on the wave-function \eqref{stationarystate}, it is simple to work out that $T$ and $R$ are as follows:
\begin{equation}
\label{transmissioncoefficients}
\begin{array}{c}
\displaystyle T_\pm = \displaystyle\frac{\left(1-\frac{D}{4}\right)\pm\, \,\mathbbm{i}\,\mbox{Im}(c_2)}{\left(1+\frac{D}{4}\right)+\frac{1}{2}\,\mathbbm{i}\,\left(\frac{c_1}{k}-k\,c_3\right)},\quad 
R_\pm = \displaystyle\frac{\mp\, \mbox{Re}(c_2) -\frac{1}{2}\,\mathbbm{i}\,\left(\frac{c_1}{k}+k\,c_3\right)}{\left(1+\frac{D}{4}\right)+\frac{1}{2}\,\mathbbm{i}\,\left(\frac{c_1}{k}-k\,c_3\right)}.
\end{array}
\end{equation}
As a result, the probability of transmission is
\begin{equation}
\label{probabilityT}
|T_+|^2=|T_-|^2=\frac{\left(1-\frac{D}{4}\right)^2+\mbox{Im}(c_2)^2}{\left(1-\frac{D}{4}\right)^2+|c_2|^2+\frac{1}{4}\left(\frac{c_1}{k}+k\,c_3\right)^2}.
\end{equation}
Recall that $D$ is the determinant of the connection matrix, i.e. $D=c_1\,c_3 + |c_2|^2$, such that $D$ is real and even under parity. The scattering coefficients for waves travelling towards the right and left are related by a parity operation (i.e. by $c_2 \to - c_2$). Clearly, the probability of transmission is unaffected by parity. If $c_2$ is real and $D=4$, no transmission takes place. Contrary to some claims in the literature, $\mbox{Im}(c_2)$ generally does affect the transmission and reflection probabilities. 

The scattering matrix $S$ is unitary for all $c_1,c_3 \in \mathbbm{R}$ and $c_2 \in \mathbbm{C}$, i.e. 
\begin{equation*}
S:=
\left(
\begin{array}{cc}
T_+ & R_-\\
R_+ & T_-
\end{array}
\right) \quad \quad \quad \mbox{ satisfies } S\,S^\dagger = \mathbbm{1},\quad\quad\quad \forall\, c_1,c_3 \in \mathbbm{R},\,c_2 \in \mathbbm{C}.
\end{equation*}
Thus probability is conserved for jump-average boundary conditions of the form \eqref{SAE6}. 

The three fundamental point interactions are quite distinct when it comes to their scattering behaviour. The transition probabilities for each of the three fundamental point interactions are as follows:
\begin{equation*}
\begin{array}{lcr}
\displaystyle |T_\pm|^2=\displaystyle\frac{1}{1 + \frac{1}{4} c_1^2 / k^2},\quad&\quad\displaystyle |T_\pm|^2=\displaystyle\frac{(c_2^2-4)\,(\overline{c}_2^2-4)}{\left(|c_2|^2+4\right)^2},\quad&\quad|T_\pm|^2=\displaystyle\frac{1}{1 +\frac{1}{4} c_3^2\, k^2}.
\end{array}
\end{equation*}
For the first/second/third fundamental point interaction, high-energy waves are more/equally/less likely to be transmitted. If $c_1=c_3=0$ and $c_2$ is purely imaginary, the probability of transmission is one. If $c_1=c_3=0$ and $c_2=\pm 2$, the probability of transmission is zero. As $c_1 \to \infty$ or $c_3 \to \infty$, the first and third point interactions become fully reflecting. If $c_2$ is real and $c_2 \to \infty$, however, the second point interaction disappears.

\section{Relation to connected SAEs}
\label{section7}

Traditionally, the literature has classified the full set of self-adjoint extensions (SAEs) as \textit{connected} or  \textit{separated}. For a given set of $c_i$, the jump-average boundary conditions fall into either one of these classes. However, the jump-average boundary conditions only form a \textit{subset} of all possible SAEs. This section and the next make these claims explicit.

Connected boundary conditions can be written in several ways, for example as (see e.g. \cite{Gesztesy1985,Seba1986,Gesztesy1987,Chernoff1993,Coutinho1997,Albeverio1998}):
\begin{equation}
\label{SAE2}
\begin{array}{c}
\left(
\begin{array}{c}
u'(0^{+})\\
u(0^{+})
\end{array}
\right) 
= 
e^{\mathbbm{i}\theta}\, 
\left(
\begin{array}{c@{\hspace{5mm}}c}
a_1 & a_2\\
a_3 & a_4
\end{array}
\right)
\,
\left(
\begin{array}{c}
u'(0^{-})\\
u(0^{-})
\end{array}
\right),\quad
\theta,a_i \in \mathbbm{R},\quad
a_1 a_4 - a_2 a_3=1.
\end{array}
\end{equation}
First, assume the jump-average boundary conditions \eqref{SAE6} hold. Then the connected parameters $a_i$ and $\theta$ can be written as a function of the jump-average parameters $c_i$ as follows:
\begin{equation}
\label{connected}
\begin{array}{lcl}
\theta = \mbox{Tan}^{-1}[1-D/4,\mbox{Im}(c_2)], &&\\
\displaystyle a_1 = \frac{D/4+1-\mbox{Re}(c_2)}{\sqrt{(D/4-1)^2+\mbox{Im}(c_2)^2}}, && \displaystyle a_2 = \frac{c_1}{\sqrt{(D/4-1)^2+\mbox{Im}(c_2)^2}},\\
\displaystyle a_3 = \frac{c_3}{\sqrt{(D/4-1)^2+\mbox{Im}(c_2)^2}}, && \displaystyle  a_4 = \frac{D/4+1+\mbox{Re}(c_2)}{\sqrt{(D/4-1)^2+\mbox{Im}(c_2)^2}}.\\
\end{array}
\end{equation}
These expressions are valid as long as $(D/4-1)^2+\mbox{Im}(c_2)^2 > 0$. Here, $\mbox{Tan}^{-1}(x,y)$ is defined so as to give the arc tangent of $y/x$, taking into account which quadrant the point $(x,y)$ is in. If $c_2$ is real, we get $\theta=0$ or $\theta=\pi$. Similar expressions can be found in \cite[p. 8]{BrascheNizhnik2013}, although the angle $\theta$ is not explicitly given there. The correspondence with that paper can be made clear by writing
\begin{equation*}
\frac{\exp\left(\,\mathbbm{i}\, \theta\right)}{\sqrt{(D/4-1)^2+\mbox{Im}(c_2)^2}}=\frac{-1}{D/4-1+\mathbbm{i}\,\mbox{Im}(c_2)}
\end{equation*}
where $\theta$ is given by \eqref{connected}. 

From the parity behaviour of the $c_i$, it follows that  $a_2$, $a_3$ and $a_1+a_4$ are even under parity, while $a_1-a_4$ is odd. The angle $\theta$, when visualized in the complex plane, is reflected in the horizontal axis under a parity operation. This implies $\theta \to - \theta$, such that $\mbox{cos}(\theta)\to \mbox{cos}(\theta)$ and $\mbox{sin}(\theta)\to -\mbox{sin}(\theta)$. 

Suppose instead that some connected boundary conditions in terms of the $a_i$ are given. The jump-average parameters $c_i$ may then be written as
\begin{equation}
\begin{array}{l}
\displaystyle c_1 = \frac{4\, a_2} {a_1 + a_4 + 2\, \mbox{cos}(\theta)},\quad c_2 = \frac{2 (-a_1 + a_4 + 2\, \mathbbm{i}\, \mbox{sin}(\theta))} {a_1 + a_4 + 2\,\mbox{cos}(\theta)}, \quad c_3 = \frac{4\, a_3} {a_1 + a_4 + 2\,\mbox{cos}(\theta)}.
\end{array}
\end{equation}
These expressions seem to be new and are valid as long as $a_1 + a_4 + 2\,\mbox{cos}(\theta)\neq 0$. We conclude that some self-adjoint extensions, namely those for which with $a_1 + a_4 + 2\,\mbox{cos}(\theta)= 0$, cannot be generated by the potential \eqref{potential}. 

\section{Relation to separated SAEs}
\label{section8}

Suppose that $(D/4-1)^2+\mbox{Im}(c_2)^2=0$. Then the jump-average boundary conditions cannot be re-written as connected boundary conditions. In this case, the jump-average boundary conditions are equivalent to boundary conditions that are traditionally known as \textit{separated}, and which can be written as
\begin{equation}
\label{SAE3}
\begin{array}{c}
\left(
\begin{array}{c}
u'(0^{+})\\
u'(0^{-})
\end{array}
\right) 
= 
\left(
\begin{array}{c@{\hspace{5mm}}c}
b^{+} & 0\\
0   & b^{-}
\end{array}
\right)
\,
\left(
\begin{array}{c}
u(0^{+})\\
u(0^{-})
\end{array}
\right),\quad b^+,b^{-} \in \mathbbm{R} \cup \infty,
\end{array}
\end{equation}
or as
\begin{equation}
\label{SAE4}
\begin{array}{c}
\left(
\begin{array}{c}
u(0^{+})\\
u(0^{-})
\end{array}
\right) 
= 
\left(
\begin{array}{c@{\hspace{5mm}}c}
\tilde{b}^+ & 0\\
0   & \tilde{b}^{-}
\end{array}
\right)
\,
\left(
\begin{array}{c}
u'(0^{+})\\
u'(0^{-})
\end{array}
\right),\quad \tilde{b}^{+},\tilde{b}^{-} \in \mathbbm{R} \cup \infty.
\end{array}
\end{equation}
In this section, we assume $D=4$ and $c_2 \in \mathbbm{R}$. This implies $c_1\,c_3 + c_2^2=4$, such that three real parameters remain, with only two degrees of freedom. It will be convenient to distinguish between three collectively exhaustive cases: $c_1$ is not zero, $c_3$ is not zero, or both $c_1$ and $c_3$ are zero:
\begin{itemize}
\item \textbf{Case 1:} $D=4$, $c_2 \in \mathbbm{R}$, and $c_1 \neq 0$. The constant $c_3$ can be eliminated since it must satisfy $c_3=(4-c_2^2)/c_1$. Then the separated parameters can be written as a function of $c_1$ and $c_2$ as follows:
\begin{equation}
\tilde{b}^+ = \frac{c_2+2}{c_1}, \quad \quad \tilde{b}^{-} = \displaystyle \frac{c_2-2}{c_1}.
\end{equation}
If, additionally, $c_2=2$ (and thus $c_3=0$), we get a Dirichlet boundary condition to the left of the origin. To the right, we get a Robin boundary condition governed by the remaining free parameter $c_1$. For $c_2=-2$, the opposite is true (Dirichlet on the right, Robin on the left). As $c_1 \to \infty$, Dirichlet boundary conditions on both sides of zero are obtained. Equivalently, the $c_i$ may be written as 
\begin{equation}
c_1 = \frac{4}{\tilde{b}^{+}-\tilde{b}^{-}}, \quad \quad c_2 = \displaystyle \frac{2(\tilde{b}^{+}+\tilde{b}^{ - })}{\tilde{b}_1-\tilde{b}^{ - }},\quad \quad c_3 = \frac{-4 \tilde{b}^{+} \tilde{b}^{ - }}{\tilde{b}^{+} - \tilde{b}^ { - }}
\end{equation}
It follows that SAEs for which $\tilde{b}^+=\tilde{b}^{ - }$ cannot be obtained using the potential \eqref{potential}. 
\item \textbf{Case 2:} $D=4$, $c_2 \in \mathbbm{R}$, and $c_3 \neq 0$. The constant $c_1$ can be eliminated, as we must have $c_1=(4 - c_2^2) / c_3$. Then the separated parameters can be written as a function of $c_2$ and $c_3$ as follows:
\begin{equation}
b^{+} = \frac{2-c_2}{c_3}, \quad \quad b^{-} = \displaystyle \frac{-2-c_2}{c_3}.
\end{equation}
If, additionally, $c_2=2$ (and thus $c_1=0$), we get a Neumann boundary condition on the right of the origin. On the left, we get a Robin boundary condition, governed by the remaining free parameter $c_3$. For $c_2=-2$, the opposite is true (Neumann on the left, Robin on the right). For $c_3 \to \infty$, Neumann conditions on both sides of zero are obtained. Equivalently, the $c_i$ may be written as 
\begin{equation}
c_1 = \frac{-4 b^{+} b^{-}}{b^{+}-b^{-}},\quad\quad c_2 = \frac{-2(b^{+}+b^{-})}{b^{+}-b^{-}},\quad \quad c_3 = \displaystyle \frac{4}{b^{+}-b^{-}},\quad \quad 
\end{equation}
As above, we find that SAEs with $b^{+}=b^{-}$ cannot be generated by the potential \eqref{potential}. 
\item \textbf{Case 3:} $D=4$, $c_2 \in \mathbbm{R}$, and $c_1=c_3=0$. This implies $c_2 = \pm 2$. If $c_2=2$, we obtain Neumann (Dirichlet) boundary conditions to the right (left) of zero. If $c_2=-2$, we obtain Dirichlet (Neumann) conditions to the right (left) of zero. 
\end{itemize}
While the jump-average boundary conditions \eqref{SAE6} do not cover \textit{all} self-adjoint extentions, they do describe those which can be generated by the potential \eqref{potential}. Having considered, in this section and the previous section, all cases using the traditional framework of connected and separated boundary conditions, the reader may appreciate the conciseness of the jump-average boundary conditions. One unanswered question, as far as we know, is whether there is a singular potential that can generate all self-adjoint extensions. 

\section{Propagator for the delta prime potential}
\label{section9}

Suppose we write down Schr{\"o}dinger's integral equation (see e.g. \cite{FeynmanHibbs1965,Schulman1981techniques,Gaveau1986,Park1996}) with a delta prime potential:
\begin{equation}
\label{SchrodingerEquation5}
\begin{array}{c}
\displaystyle\psi(y,t|x,s)=\displaystyle \psi_0(y,t|x,s)-c\,\mathbbm{i} \,\int_s^t d\tau \int^\infty_{-\infty} d\alpha\;\psi_0(y,t|\alpha,\tau)\,\delta'(\alpha)\,\psi(\alpha ,\tau |x,s),\\
\end{array}
\end{equation}
where $c \in \mathbbm{R}$, the delta function was defined in \eqref{Dirac1}, $\psi_0$ was defined in \eqref{freepropagator}, and it is assumed that $\psi(\cdot,t|x,s)\in W_2^2(\mathbbm{R}\backslash\{0\})$. As highlighted in section \ref{section4}, the integral equation allows no solutions if the definition of the delta prime is such that $\int \delta' \psi$ blows up for discontinuous $\psi$. With our distributional definition \eqref{Dirac1}, however, the integral equation can be solved in closed form as follows:
\begin{equation}
\label{deltaprimeSsolution}
\psi(y,t|x,s)
=\psi_0(y,t|x,s)\;+\left\{
\begin{array}{l@{\hspace{2mm}}r@{\hspace{10mm}}l}
\displaystyle\;+\;\frac{4c }{4+c ^2} &\psi_0(y,t|-x,s),&x>0,y>0,\\[2ex]
\displaystyle\;-\;\frac{2c ^2}{4+c ^2}&\psi_0(y,t|x,s),&x>0,y<0,\\[2ex]
\displaystyle\;-\;\frac{2c ^2}{4+c ^2}&\psi_0(y,t|x,s),&x<0,y>0,\\[2ex]
\displaystyle\;-\;\frac{4c }{4+c ^2}&\psi_0(y,t|-x,s),&x<0,y<0.
\end{array}
\right.
\end{equation}
As far as we are aware, this exact solution to Schr{\"o}dinger's equation is new. What's more, it is remarably simple; much simpler, in fact, than the well-known propagator for the delta potential. The calculation is carried out below. From the explicit solution, we can verify that the propagator $\psi$ satisfies the following boundary conditions:
\begin{equation}
\left(
\begin{array}{c}
[\psi'(0,t|x,s)]\\
{[\psi(0,t|x,s)]}
\end{array}
\right)
= 
\left(
\begin{array}{c@{\hspace{5mm}}c}
0& -c \\
c & 0
\end{array}
\right)
\left(
\begin{array}{c}
\{\psi(0,t|x,s)\}\\
\{\psi'(0,t|x,s)\}
\end{array}
\right).
\end{equation}
These boundary conditions are of the jump-average form \eqref{SAE6}, with $c_2=c\in \mathbbm{R}$ and $c_1=c_3=0$. The derived boundary conditions are consistent with the independently derived boundary conditions in \cite{KurasovElander1993,Kurasov1994,Albeverio1998,gadella2009bound}. Interestingly, those derivations were based on Schr{\"o}dinger's differential (rather than integral) equation.

As can be seen from the solution, $c=\pm 2$ implies that the propagator is zero for $x$ and $y$ on opposite sides of the origin. For $c=2$, the propagator satisfies Neumann boundary conditions at $0^+$ and Dirichlet boundary conditions at $0^-$. The opposite is true for $c =-2$. If we focus on $0^{+}$, we have Dirichlet (Neumann) boundary conditions for $c=-2$ ($c=+2$). As in section \ref{section5}, the only difference between Dirichlet and Neumann boundary conditions on a given side of the boundary resides in the \textit{sign} of the delta prime potential (see also \cite{lange2012}). 

For $c\neq\pm 2$, the potential is partially transparent with the scattering matrix given in section \ref{section6}. Recently, several authors have found the delta prime potential to be transparent only for particular values of $c $; see \cite{christiansen2003existence,Coutinho1997,Coutinho2005,zolotaryuk2006scattering,Toyama2007,zolotaryuk2007resonant,zolotaryuk2010boundary}. The difference is attributable to the construction of the delta prime function. Here it is expressly defined so as to be compatible with discontinuous test functions. For methods 2-4 in Table \ref{Table1}, the integral equation \eqref{SchrodingerEquation5} would not exist for discontinuous $\psi$.

The solution to Schr{\"o}dinger's integral equation was obtained as follows. By repeatedly substituting the integral equation \eqref{SchrodingerEquation5} into itself, the solution may be written as:
\begin{equation*}
\label{seriessolution}
\begin{array}{lr@{\hspace{1mm}}c@{\hspace{1mm}}}
\displaystyle\psi(y,t|x,s)&=&\displaystyle \psi_0(y,t|x,s)+\overset{\infty}{\underset{i=1}\sum} (-1)^i\,\psi_i(y,t|x,s),
\end{array}
\end{equation*}
where the correction terms $\psi_i$ are defined recursively as
\begin{equation*}
\begin{array}{lr@{\hspace{1mm}}c@{\hspace{1mm}}}
\displaystyle\psi_i(y,t|x,s)&=&\displaystyle\mathbbm{i}\,\int_s^t d\tau \int_{-\infty}^{\infty} d\alpha\;\psi_0(y,t|\alpha,\tau)\,V(\alpha)\,\psi_{i-1}(\alpha ,\tau |x,s),
\end{array}
\end{equation*}
and $V(x)=c\,\delta'(x)$. For singular potentials, the recursive structure of the correction terms should be carefully observed, i.e.
\begin{equation*}
\begin{array}{lr@{\hspace{1mm}}c@{\hspace{1mm}}}
\displaystyle\psi_2&=&\displaystyle \int \int \psi_0\,V\,\int \int \psi_0\,V\,\psi_0\neq\displaystyle \int \int \int \int \psi_0\,V\,\psi_0\,V\,\psi_0.
\end{array}
\end{equation*}
In other words, the interchange of integrals and distributions is not generally allowed, i.e. integrals cannot be pulled to the front. 

The first correction term is $\psi_1=c\mathbbm{i}\int \int \psi_0 \delta' \psi_0$. Since $\psi_0$ is continuously differentiable across zero, no ambiguities whatsover arise regarding the interpretation of the $\delta'$-function. Performing the integration, we obtain the following expression:
\begin{equation*}
\makebox[12cm][l]
{
$
\psi_1(y,t|x,s)
=\left\{
\begin{array}{lcr}
-c\, \psi_0(y,t|-x,s)&&x>0,y>0,\\
0&&x>0,y<0,\\
0&&x<0,y>0,\\
c\, \psi_0(y,t|-x,s)&&x<0,y<0.
\end{array}
\right.
$
}
\end{equation*}
Since $\psi_1$ is discontinuous, the exact distributional definition of the delta prime is crucial for the calculation of $\psi_2=c\mathbbm{i}\int\int \psi_0 \delta' \psi_1$. Using our definition of the delta function, all correction terms are finite and can be calculated explicitly. For e.g. $\psi_2$, $\psi_3$, $\psi_4$ and $\psi_5$, we obtain the following expressions:
\begin{equation*}
\makebox[12cm][l]
{
$
\psi_2(y,t|x,s)
=\left\{
\begin{array}{lcr}
0&&x>0,y>0,\\
-\frac{1}{2}c^2 \psi_0(y,t|x,s)&&x>0,y<0,\\
-\frac{1}{2}c^2 \psi_0(y,t|x,s)&&x<0,y>0,\\
0&&x<0,y<0.
\end{array}
\right.
$
}
\end{equation*}
\begin{equation*}
\makebox[12cm][l]
{
$
\psi_3(y,t|x,s)
=\left\{
\begin{array}{lcr}
\frac{1}{4}c^3 \,\psi_0(y,t|-x,s)&&x>0,y>0,\\
0&&x>0,y<0,\\
0&&x<0,y>0,\\
-\frac{1}{4}c^3 \,\psi_0(y,t|-x,s)&&x<0,y<0.
\end{array}
\right.
$
}
\end{equation*}
\begin{equation*}
\makebox[12cm][l]
{
$
\psi_4(y,t|x,s)
=\left\{
\begin{array}{lcr}
0&&x>0,y>0,\\
\frac{1}{2^3}c^4\,\psi_0(y,t|x,s)&&x>0,y<0,\\
\frac{1}{2^3}c^4 \,\psi_0(y,t|x,s)&&x<0,y>0,\\
0&&x<0,y<0.
\end{array}
\right.
$
}
\end{equation*}
\begin{equation*}
\makebox[12cm][l]
{
$
\psi_5(y,t|x,s)
=\left\{
\begin{array}{lcr}
-\frac{1}{2^4}c^5 \,\psi_0(y,t|-x,s)&&x>0,y>0,\\
0&&x>0,y<0,\\
0&&x<0,y>0,\\
\frac{1}{2^4}c^5  \,\psi_0(y,t|-x,s)&&x<0,y<0.
\end{array}
\right.
$
}
\end{equation*}
It becomes clear that the following series solution arises:
\begin{equation*}
\psi(y,t|x,s)
=\left\{
\begin{array}{lcr}
\psi_0(y,t,x,s) + 2 \Big(\frac{c}{2}-\frac{c^3}{2^3}+\frac{c^5}{2^5}-\frac{c^7}{2^7}+\ldots   \Big)\,\psi_0(y,t|-x,s)&&x>0,y>0,\\[2ex]
\psi_0(y,t,x,s) - 2 \Big(\frac{c^2}{2^2}-\frac{c^4}{2^4}+\frac{c^6}{2^6}-\frac{c^8}{2^8}+\ldots   \Big)\,\psi_0(y,t|x,s)&&x>0,y<0,\\[2ex]
\psi_0(y,t,x,s) - 2 \Big(\frac{c^2}{2^2}-\frac{c^4}{2^4}+\frac{c^6}{2^6}-\frac{c^8}{2^8}+\ldots   \Big)\,\psi_0(y,t|x,s)&&x<0,y>0,\\[2ex]
\psi_0(y,t,x,s) - 2 \Big(\frac{c}{2}-\frac{c^3}{2^3}+\frac{c^5}{2^5}-\frac{c^7}{2^7}+\ldots   \Big)\,\psi_0(y,t|-x,s)&&x<0,y<0.
\end{array}
\right.
\end{equation*}
These expressions may be recognised as the Taylor series expansions of the exact solution \eqref{deltaprimeSsolution}. Although the solution was derived using the series expansion, it can be verified directly that the proposed solution satisfies the integral equation.

\section{Super-singular potentials}
\label{section10}

This section considers Schr{\"o}dinger's integral equation with the potential $V(x)=c\,\delta^{(n)}(x)$ with $c \in \mathbbm{R}$ for $n \geq 1$. These super-singular potentials are interesting as they cannot be incorporated into the normal framework of self-adjoint extensions. We show that the resulting boundary conditions are of the jump-average form \eqref{SAE6}, with the crucial difference that the constants $c_i$ depend on the energy $E=k^2$. Specifically, we show that
\begin{equation}
\label{evenodd}
V(x) = c\,\delta^{(n)}(x)
\Rightarrow
\left\{
\begin{array}{l}
\mbox{For even}\,n\mbox{, BCs \eqref{SAE6} with:} \\
c_1(k) = c\, 2^{n-1} (\mathbbm{i} k)^{n}, \\
c_2(k) = 0,\\
c_3(k) = -c\, 2^{n-1} (\mathbbm{i} k)^{n-2},
\end{array}
\right.
\quad\quad
\begin{array}{l}
\mbox{For odd}\, n\mbox{, BCs \eqref{SAE6} with:}\\
c_1(k) = 0\\
c_2(k) = c\, (2\mathbbm{i} k)^{n-1},\\
c_3(k) = 0 .
\end{array}
\end{equation}
For even $n$, only even constants $c_i$ are non-zero (i.e. $c_1$ and $c_3$). For odd $n$, only the odd constant $c_2$ is non-zero. We also note that all $c_i$ are real. Setting $n$ equal to 1 yields the boundary conditions of section \ref{section9}. For $n>1$, however, we obtain boundary conditions that are \textit{energy-dependent} in the sense that the $c_i$'s depend on $k$. For $n=2$, for example, only $c_1$ depends on the energy $E$. These boundary conditions seem to be new. 

The transmission and reflection coefficients are given by \eqref{transmissioncoefficients} with the $c_i$ as above. Probability is conserved for all boundary conditions of the jump-average form, even if the constants $c_i$ depend on $k$. Crucially, therefore, probablity is conserved. 

Jump-average boundary conditions of the form \eqref{SAE6} are self-adjoint when the parameters $c_i$ are constant, but not when they depend on the energy through $k$. Thus it appears that these super-singular interactions conserve probability without being self-adjoint. 

The proof of \eqref{evenodd} consists of two steps. First, we show that $\psi(\cdot,t|x,s)$ must satisfy 
\begin{equation}
\label{generalBCs}
\begin{array}{r@{\hspace{1mm}}c@{\hspace{1mm}}ll}
\displaystyle [\psi^{(1)}(0,t|x,s)]&=&\displaystyle c \,2^{n-1}\,(-1)^{n}&\{\psi^{(n)}(0,t|x,s)\},\\
\displaystyle [\psi(0,t|x,s)]&=&\displaystyle c \,2^{n-1}\,(-1)^{(n-1)}&\{\psi^{(n-1)}(0,t|x,s)\}.
\end{array}
\end{equation}
These boundary conditions are slightly different from the so-called Griffiths' boundary conditions \cite{Griffiths1993}, as the numerical prefactors on the right-hand side are different. While the original derivation of Griffiths' boundary conditions is known to be erroneous \cite{Coutinho1997,Coutinho2004,Coutinho2005}, the resulting boundary conditions have generated interest in their own right (see e.g. \cite{coutinho2012one}). The second step of the proof shows that the boundary conditions \eqref{generalBCs} are equivalent to \eqref{evenodd}. 

We now turn to the first step of the proof. We begin by extending lemma \eqref{fconditions} to state that all \textit{odd} derivatives of the auxiliary function $f$, defind as
\begin{equation}
\begin{array}{lr@{\hspace{1mm}}c@{\hspace{1mm}}}
f(y,t|x,s):=\displaystyle -\mathbbm{i}\int_s^t \; \psi_0(y,t|0,\tau)\,g(\tau|x,s)\,d\tau,
\end{array}
\end{equation}
are discontinuous as follows:
\begin{equation}
\label{fconditions2}
\displaystyle [f^{(k)}(0,t|x,s)]
=
\left\{
\begin{array}{l@{\hspace{5mm}}l}
0& k \in \mbox{0, even,}\\[2ex]
\displaystyle\,\left(-\mathbbm{i}\frac{d}{dt}\right)^{\frac{k-1}{2}} g(t|x,s)& k \in \mbox{ odd.}\\[2ex]
\end{array}
\right.
\end{equation} 
This will be useful in the context of Schr{\"o}dinger's integral equation, which can be re-written for $V(x)=c\, \delta^{(n)}(x)$ as follows:
\begin{equation*}
\begin{array}{rcl}
\displaystyle\psi(y,t|x,s)&=&\psi_0(y,t|x,s)-c\,\mathbbm{i}\displaystyle \int_s^t d\tau \displaystyle \int_{-\infty}^{\infty} d\alpha\;\psi_0(y,t|\alpha,\tau)\,\delta^{(n)}(\alpha)\,\psi(\alpha ,\tau |x,s),\\[3ex]
&=&\displaystyle\psi_0(y,t|x,s)-c\,\mathbbm{i}\,
(-1)^n
\underset{i=0}{\overset{n}{\sum}} 
\left(
\begin{array}{c}
n\\
i
\end{array}
\right)
\displaystyle\int_s^t 
\psi_0^{(i)}(y,t|0,\tau)\;\{\psi^{(n-i)}(0,\tau |x,s) \}\,d\tau,\\[3ex]
&=&\displaystyle\psi_0(y,t|x,s)-c\,\mathbbm{i}\, \, 
(-1)^{n}
\underset{i=0}{\overset{n}{\sum}} 

\left(
\begin{array}{c}
n\\
i
\end{array}
\right)
(-1)^{i}
\frac{d^{i}}{dy^{i}}
\int_s^t
\displaystyle

\;\psi_0(y,t|0,\tau)\;\{\psi^{(n-i)}(0,\tau |x,s) \}\,d\tau.\\[2ex]
\end{array}
\end{equation*}
The second line used the definition of the delta function \eqref{Dirac1}, as well as the fact that the free propagator can be pulled out of the averaging operator. The third line used the fact that $d/d\alpha\, \psi_0(y,t|\alpha,\tau) = - d/dy\, \psi_0(y,t|\alpha,\tau)$. Further, derivatives $d/dy$ can be pulled out of the integral sign for all $y\neq 0$, since $\psi(\cdot,t|x,s)$ is smooth away from the origin.

Referring to \eqref{fconditions2}, we realise that integral terms with an \textit{odd} number of derivatives $(d/dy)^{i}$ are discontinuous. Equally, integral terms with an \textit{even} number of derivatives $(d/dy)^{i}$ are continuous, but then the first derivative with respect to $y$ is discontinuous. Thus the discontinuity in the value of $\psi(\cdot,t|x,s)$ is determined by the sum over all odd $i$, while the discontinuity in the derivative is determined by the sum over all even $i$. Using \eqref{fconditions2}, we get
\begin{equation*}
\begin{array}{r@{\hspace{1mm}}c@{\hspace{1mm}}rl}
\displaystyle [\psi^{(1)}(0,t|x,s)]&=&\displaystyle c\,(-1)^n&\displaystyle\underset{i=0,2,4\ldots\leq n}{\sum}
\left(
\begin{array}{c}
n\\
i
\end{array}
\right)
\left(-\mathbbm{i}\frac{d}{dt}\right)^{\frac{i}{2}}\{\psi^{(n-i)}(0,t|x,s)\},\\
\displaystyle [\psi(0,t|x,s)]&=&-\displaystyle c\,(-1)^n &\displaystyle
\underset{i=1,3,5\ldots \leq n}{\sum} 
\left(
\begin{array}{c}
n\\
i
\end{array}
\right)
\left(-\mathbbm{i}\frac{d}{dt}\right)^{\frac{i-1}{2}}\{\psi^{(n-i)}(0,t|x,s)\}.
\end{array}
\end{equation*}
Above and below zero, $\psi(\cdot,\cdot|x,s)$ satisfies the free Schr{\"o}dinger equation. As a result, we have  $-\mathbbm{i}\partial_t\{\psi^{(n)}(0,t|x,s)\}=\{\psi^{(n+2)}(0,t|x,s)\}$ and thus
\begin{equation*}
\begin{array}{r@{\hspace{1mm}}c@{\hspace{1mm}}rl}
\displaystyle [\psi^{(1)}(0,t|x,s)]&=&\displaystyle c\,(-1)^n&\displaystyle
\underset{i=0,2,4\ldots\leq n}{\sum}
\left(
\begin{array}{c}
n\\
i
\end{array}
\right)
\{\psi^{(n)}(0,t|x,s)\},\\
\displaystyle [\psi(0,t|x,s)]&=&-\displaystyle c\,(-1)^n &\displaystyle
\underset{i=1,3,5\ldots \leq n}{\sum}  
\left(
\begin{array}{c}
n\\
i
\end{array}
\right)
\{\psi^{(n-1)}(0,t|x,s)\}.
\end{array}
\end{equation*}
As claimed, this implies
\begin{equation}
\label{deltaderivativeBCs}
\begin{array}{r@{\hspace{1mm}}c@{\hspace{1mm}}ll}
\displaystyle [\psi^{(1)}(0,t|x,s)]&=&\displaystyle c \,2^{n-1}\,(-1)^{n}&\{\psi^{(n)}(0,t|x,s)\},\\
\displaystyle [\psi(0,t|x,s)]&=&\displaystyle c \,2^{n-1}\,(-1)^{(n-1)}&\{\psi^{(n-1)}(0,t|x,s)\},
\end{array}
\end{equation}
where the combinatioral factors on the right-hand side arise from the summation over half of all the binomial coefficients, i.e. $2^n/2=2^{n-1}$. 
 
Moving on to the second step, we will consider separately even and odd $n$. Suppose $n$ is even and consider a stationary state $\psi$ of the form \eqref{stationarystate}. Then $\{\psi^{(n)}(0)\}=(\mathbbm{i} k)^{n}\{\psi(0)\}$ and $\{\psi^{(n-1)}(0)\}=(\mathbbm{i} k)^{n-2}\{\psi^{(1)}(0)\}$, such that the boundary conditions \eqref{deltaderivativeBCs} can be written as follows:
\begin{equation}
\left(
\begin{array}{c}
[\psi'(0)]\\
{[\psi(0)]}
\end{array}
\right)
=\;c\,2^{n-1}
\left(
\begin{array}{c@{\hspace{7mm}}c}
(\mathbbm{i} k)^{n}& 0\\
0 & -(\mathbbm{i} k)^{n-2}
\end{array}
\right)
\left(
\begin{array}{c}
\{\psi(0)\}\\
\{\psi'(0)\}
\end{array}
\right).
\end{equation}
These boundary conditions are of the jump-average form \eqref{SAE6}, with $c_1$ and $c_3$ as in \eqref{evenodd}. The scattering matrix is unitary for all boundary conditions of the jump-average form, and thus probability is conserved. 

Now suppose $n$ is odd, such that $\{\psi^{(n)}(0)\}=(\mathbbm{i} k)^{n-1}\{\psi^{(1)}(0)\}$ and $\{\psi^{(n-1)}(0)\}=(\mathbbm{i} k)^{n-1}\{\psi(0)\}$. Then the boundary conditions \eqref{deltaderivativeBCs} can be written as follows:
\begin{equation}
\left(
\begin{array}{c}
[\psi'(0)]\\
{[\psi(0)]}
\end{array}
\right)
=c\,2^{n-1}
\left(
\begin{array}{c@{\hspace{7mm}}c}
0& -\,(\mathbbm{i} k)^{n-1}\\
\,(\mathbbm{i} k)^{n-1} & 0
\end{array}
\right)
\left(
\begin{array}{c}
\{\psi(0)\}\\
\{\psi'(0)\}
\end{array}
\right).
\end{equation}
Again, these boundary conditions are of the jump-average form \eqref{SAE6}, with $c_2$ as in \eqref{evenodd}. 

We conclude that the potential $\delta^{(n)}$ is permissable if probability conservation is imposed. Despite being of self-adjoint \textit{form}, however, the boundary conditions are not self-adjoint since the parameters depend on the energy $k^2$. We leave open the question whether the derived boundary conditions can be made self-adjoint by considering, in addition to the real line, some internal space at the origin. 

\section{Conclusion}
\label{section11}

This paper considered the integral form of Schr{\"o}dinger's equation, where the potential is given by a distribution that is defined on the space of discontinuous functions. Broadly, it has shown that Schr{\"o}dinger's integral equation is a viable tool for studying singular interactions in quantum mechanics. 

Section \ref{section2} re-derived Kurasov's based purely on symmetry considerations. Section \ref{section3} showed that the associated boundary conditions can be expressed quite naturally using the jump-average representation. 

Section \ref{section4} showed that the same result can be obtained relatively simply in the context of Schr{\"o}dinger's integral equation. This result hints at a deeper equivalence between both approaches, which are normally thought to be equivalent only for smooth potentials. 

Section \ref{section5} proposed an extension of Kurasov's result to hypersurfaces. Our result is based on an informal treatment only, but points at an interesting connection with classical potential theory. It turns out that the surface delta prime potential can generate solutions to the Dirichlet and Neumann boundary value problems, where the only difference between these two classical problems resides in the sign of the potential. 

Section \ref{section6} derived the scattering matrix in one dimension, and showed that for the first, second and third fundamental point interactions, high-energy waves are more, equally and less likely to be transmitted, respectively.

Sections \ref{section7} and \ref{section8}  showed that the jump-average boundary conditions form a \textit{subset} of all possible connected and separated self-adjoint extensions. Whether a singular potential exists that can generate all SAEs remains an open question.

Section \ref{section9} solved Schr{\"o}dinger's integral equation for the delta prime potential.  While the propagator for the delta potential has long been known, the propagator for the delta prime potential derived here is new. Our solution suffers from none of the drawbacks often found in the literature, such as an ambiguous Schr{\"o}dinger equation, arbitrary boundary conditions or ambiguous limits. By confronting the issue of a discontinuous solution head on, all ambiguities disappear. In contrast with some recent findings, the delta prime potential turns out to be partially transparent for almost all values of the coupling constant. 

Section \ref{section10} used the same method to derive boundary conditions for higher-order derivatives of the delta potential. It turns out that the boundary conditions associated with these super-singular potentials are of the jump-average form --- but with the crucial difference that the parameters depend on the energy of the solution. While probability is conserved, these energy-dependent boundary conditions are not self-adjoint when considering only the real line. If we consider a larger space, containing some internal space at the origin, then it is possible that the derived boundary conditions are, in fact, self-adjoint. This may be an interesting avenue for further research.



\bibliographystyle{JHEP2}
\bibliography{bib}


\end{document}